\documentclass[review]{elsarticle}

\usepackage{lineno,hyperref}
\modulolinenumbers[5]
\usepackage{graphicx}
\usepackage{dcolumn}
\usepackage{bm}
\usepackage[T1]{fontenc} 
\usepackage{booktabs} 
\usepackage{ dsfont }
\usepackage[utf8]{inputenc}
\usepackage[T1]{fontenc}
\usepackage{mathptmx}
\usepackage{amsmath}

\usepackage[T1]{fontenc} 
\usepackage{booktabs} 
\usepackage{ dsfont }
\usepackage{float}
\usepackage{dsfont}

\usepackage{booktabs}
\usepackage{multirow}
\usepackage{graphicx}
\usepackage{dcolumn}
\usepackage{bm}
\usepackage[T1]{fontenc} 
\usepackage{booktabs} 
\usepackage{ dsfont }
\usepackage{amsmath, amsthm, amssymb, amsfonts}
\usepackage{xcolor}

\journal{New Astronomy}

\begin{document}


\title{Feasibility of Cosmic Microwave Background Observations Using Radiometers Based on Whispering Gallery Mode Resonators.}

\author[label1,label2]{Javier De Miguel-Hern\'andez\corref{cor1}}
\address[label1]{Instituto de Astrof\'isica de Canarias, E-38200 La Laguna, Tenerife, Spain\\}
\address[label2]{Departamento de Astrof\'isica, Universidad de La Laguna, E-38206 La Laguna, Tenerife, Spain}

\cortext[cor1]{Corresponding author}

\ead{jmiguel@iac.es}

\author[label1,label2]{Roger J. Hoyland}

\begin{abstract}
The fundamentals of the whispering gallery mode (WGM) resonators are well established in the literature, with several successful proof-of-concept experiments. One remarkable benefit of this technology is the room-temperature operation. This characteristic could be used to build a new generation of radiometers that do not need to be cooled down to cryogenic temperatures to reach high sensitivities. In this article, a study of the viability of technological transfer is undertaken, beginning with a brief review of the theoretical background that will be applied and leading to a proposal for a novel spectro-polarimeter design. Simulations for a radiometer based on WGM resonance are analyzed and compared with state-of-the-art coherent receivers. The results are then discussed, finding that although promising, WGM technology needs R\&D in several directions in order to be competitive, whose are suggested by the authors with the idea of inspiring the work of the researchers in the field towards a new direction or approach.
\begin{description}
\item[Keywords]
CMB experiments, CMB detectors, CMB polarisation, photonics, astrophysics.
\end{description}
\end{abstract}

\maketitle


\section{Introduction}
\label{sec:Introduction}
The latest advances in experiments and theory on the mechanism of up-conversion of THz signals are quite remarkable (e.g., \cite{Takida}, \cite{Takida2}, \cite{Yao}). The fundamentals of the whispering gallery mode (WGM) resonators are well established as well. There are several proof-of-concept experiments demonstrating the principle of operation of the parametric up-conversion for microwave signals at THz frequencies using WGM resonators ({\it e.g.} \cite{Sedlmeir:16}, \cite{Rueda:16}, \cite{Khan:07}, \cite{Strekalov:09}, \cite{Strekalov:09-2}). 
In the setup of these experiments, a waveguide and a diamond prism are used to couple microwaves and optical waves from a laser pump to a resonator ring, and a photon-counter avalanche photodiodes (APDs) are commonly used to detect the \textit{up-converted} photons.
On the other hand, standard radiometers need to be cooled down to cryogenic temperatures in order to achieve high sensitivities. However, WGM resonators and photon-counters can theoretically operate at room temperature achieving high sensitivities if the efficiency of conversion is high enough. Because of this characteristic, it is interesting to study the viability of a technological transfer to radio-astronomy of these devices, looking at the first prototypes from a purely theoretical point of view. This is the aim of this manuscript, focusing on Cosmic Microwave Background (CMB) observations but exportable to other purposes.

In this article the general issues of the application of the WGM based detectors to radio-astronomy are discussed, obtaining several remarkable conclusions. In section \ref{sec:Background}, the theoretical framework of the WGM resonators is briefly presented. In section \ref{sec:Sensitivity}, the more simple radiometer based on WGM resonators technology is described, calculating its theoretical sensitivity and comparing with High Electron Mobility Transistor (HEMT) and microwave kinetic inductance detector (MKID) radiometers.\footnote{We avoid a direct comparison between classic bolometer, transition-edge sensor (TES) or similar and WGM radiometers, since these cryogenic detectors are incoherent, while HEMP, MKID and WGM maintain phase coherence.} In section \ref{sec:Radiometer} a potential design of microwave spectro-polarimeter based on WGM resonators is presented. Finally, conclusions and future lines of work are discussed in section \ref{sec:Discussion}.

It is important to emphasize that the study in this document is theoretical, on an ideal system free of systematic errors (fluctuations of gain and temperature, losses, ...). Here we present a theoretical and objective study of the feasibility of using WGM based radiometers in the mid-term future, when the state-of-the-art has advanced sufficiently, if this happens. We clarify as well that we are not involved in research to improve WGM resonators, and that our only interest is their applications to observational cosmology.

\section{Background}
\label{sec:Background}

\begin{figure}[ht!]
		\includegraphics[width=0.7\textwidth]{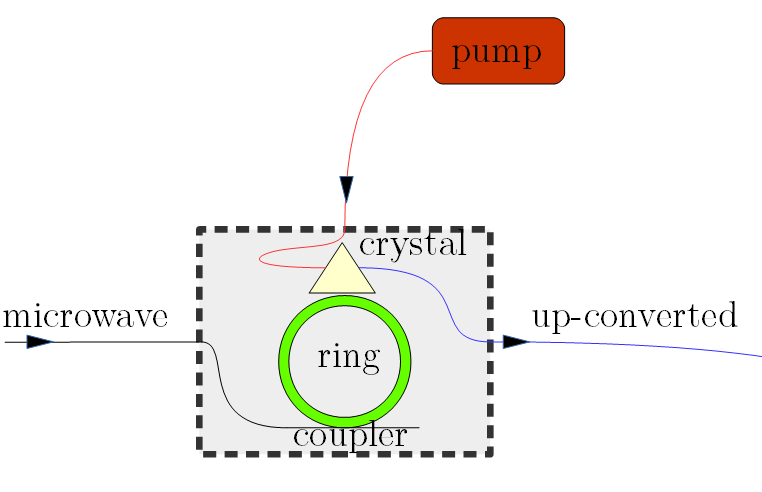}
		\centering 
		\caption{Schematic of a WGM resonator.}
		\label{fig5_0}
	\end{figure}

The theoretical background of this issue is very deep and rich in detail  (e.g., see \cite{mandel_wolf_1995}, \cite{Boyd}, \cite{Florian}, \cite{Matsko:07}, \cite{PhysRevA.77.043812}). However, it is not necessary to keep in mind each one of these details in order to correctly follow the work presented in this article. In this brief section we focus only on these points that are relevant in order to advance through this manuscript, while a more extended derivation is included in Appendix A.\newline \newline
A schematic diagram of a WGM resonator is shown in Fig. \ref{fig5_0}, where a microwave input (black) excites a resonator (ring) through a coupler, while the laser pump (red) is coupled to the ring through a diamond prism (crystal). Photon annihilation and side-band creation is produced within the resonator ring and the output, including the up-converted photons, is recovered from the crystal (blue) in direction to an APD.

\begin{figure}[ht!]
		\includegraphics[width=0.7\textwidth]{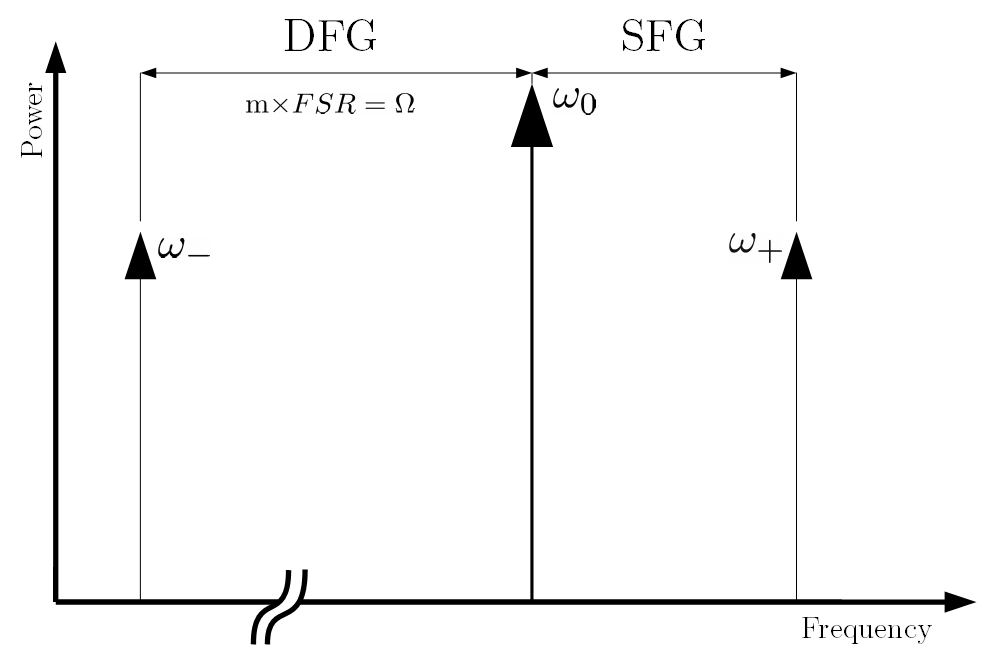}
		\centering 
		\caption{Up-conversion scheme. Stokes (to higher frequency) and anti-Stokes (to lower-frequency) side-bands are generated. Difference-frequency generation (DFG) and sum-frequency generation (SFG) are represented.}
		\label{fig3_1}
	\end{figure}

The production of side-bands during the annihilation of photons is represented in Fig. \ref{fig3_1}, where $\Omega=\omega-\omega_-$ is the microwave pulse, $\omega_0$ is the optical pump pulse and the converted frequency is given by $\omega_+=\omega+\Omega$. The factor $m$ is an integer and $FSR=1/\tau$, being $\tau \simeq 2 \pi R n/c$ the {\it round trip time}, with $R$ the resonator radius, $n$ the refraction index and $c$ the speed of light.

Thus, within the resonator annihilation of photons producing side-bands exist, with \textit{selection rules} given by photon energy and momentum conservation (see Appendix A). 
An important feature is the \textit{effective temperature} ($T_e$), since it establishes that the temperature of the system can be naturally modified trough the coupling rate. This is,

\begin{equation}
T_{\text{e}}=\frac{\gamma}{\gamma+\gamma'}T_{\text{s}}+\frac{\gamma'}{\gamma+\gamma'}T_{\text{i}}
\;,
\label{equation_3019}
\end{equation}

where addends for both the surrounding (s) medium and the resonator material (i) are included, and the coupling rate is

\begin{equation} 
\gamma=\frac{| \mathfrak{t} |^2}{2\tau}
\;,
\label{equation_3016}
\end{equation}

with $\mathfrak{t}\in \mathds{C}$ being the transmission coefficient, which can be derived from the Fresnel's equations, while the field-loss rate is given by

\begin{equation}
\gamma'=\frac{1-a}{\tau}
\;,
\label{equation_3017}
\end{equation} \newline

where $a$ is the attenuation factor. 

The situation $\gamma<\gamma'$ is commonly called the \textit{undercoupling} case, $\gamma>\gamma'$ \textit{overcoupling} and $\gamma = \gamma '$ is the \textit{critical coupling} case. From Eq. (\ref{equation_3019}) it can derived that if the environment is {\it cold} and and the resonator is over-coupled, $T_{\text{e}}$ can be decreased. This property would be relevant for astronomical purposes, because it implies that the thermal noise can be naturally decreased. \newline

\section{Detector Unit Analysis}
\label{sec:Sensitivity}
\begin{figure*}[ht!]
\centering
		\includegraphics[width=1\textwidth]{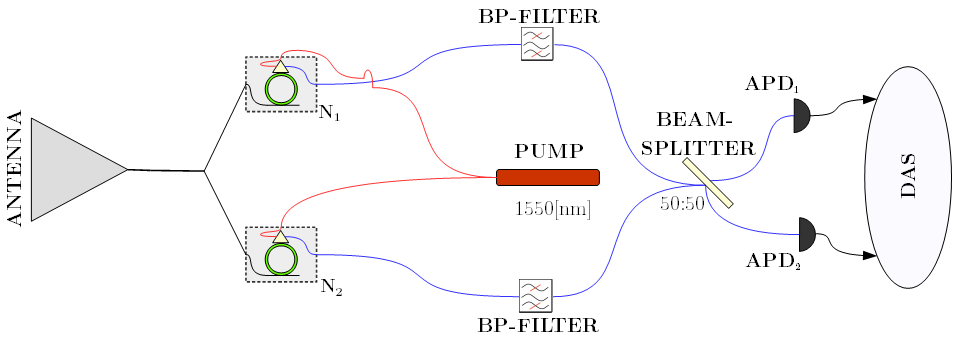}
		\centering 
		\caption{Conceptual design of a detector unit based on WGM resonators technology.}
		\label{fig3_2}
	\end{figure*}

The analysis of the sensitivity of the heterodyne detector unit of Fig. \ref{fig3_2} is presented in the following lines. In this figure, the signal captured by a microwave antenna is divided and sent to two twin WGM-resonator units, who share an optical solid-state laser. The up-converted output signals are sent to a beam-splitter the outputs of which go to two APDs before being sent to the data acquisition system (DAS). The up-converted photons and the laser-pump maintain phase coherence, so a pseudo cross-correlation technique can be applied. Hence, the signal amplitude is the difference between the two APDs and the mitigation of low frequency gain fluctuations is possible. This technique is commonly used to ameliorate problems associated with low frequency gain and noise temperature fluctuations in detector systems. These fluctuations limit the duration of the data integration that is used to increase sensitivity. We have explicitly stated that this is a theoretical study of sensitivity of a novel technology. These studies are always on ideal systems, which do not present systematic errors such as gain and temperature fluctuations. This calculation also ignores the loss in  the components used to couple the input signal to the WGM resonator, which can be substantial when the components are maintained at room temperature. However, note that some of these systematics are indirectly expressed through $\eta_+$, which is an experimental magnitude.
Note that in order to allow absolute CMB measurement, a twin radiometer pointing to another matched and stable microwave source should be added to this schema, and both outputs compared. However, the idea of this section is to show the simplest system, where a relative calibration against a well-known source (e.g., a ground-based source) is considered. Note that the conclusions are valid for both absolute or relative calibration. \newline

It's well known that a phenomena is in the quantum region when $A[\frac {kg \, m^2}{s}]\simeq \hbar[J \cdot s]$, where $A$ denotes "action". This idea applied to the case of a radiometer yields

\begin{equation}
\frac{Power}{2\pi\nu^2}\leqslant h\, \;.
\label{equation_3047}
\end{equation}
In (\ref{equation_3047}) $h$ is the Planck's constant. This is the reason why temperatures $T_{sys}\sim 20$ K and frequencies over 70 GHz present a non-negligible quantum-noise contribution. Therefore, the uncertainty in the measurement of the temperature ($\delta T$) and so the sensitivity of an ideal system should be considered by the addition of at least three independent terms 

\begin{equation}
\delta T_{RMS}=\delta T_{th}+\delta T_{shot}+\delta T_{SP}\, \;,
\label{equation_3048}
\end{equation}

where $\delta T_{th}$ is the thermal noise contribution, $\delta T_{shot}$ is the shot-noise contribution and $\delta T_{SP}$ is the statistical uncertainty of the incoming CMB photons, given by a Super-Poissionan distribution \cite{Fox}\footnote{Even though the assumption of a Poisson distribution is used extensively, this distribution is, in principle, a better approximation.}.

The first term in (\ref{equation_3048}), $\delta T_{th}$, is given by

\begin{equation}
\delta T_{th}=\frac{h \nu_{CMB}} {k_B \, ln \left(\frac{\eta_+ \, \Delta \nu \, QE}{N_{xx}}+1\right)}
\, \;,
\label{equation_3049}
\end{equation}

where $\nu_{CMB}$ is the frequency associated to the CMB photons, $\eta_+$ is the up-conversion efficiency, $\Delta \nu$ is the bandwidth, $QE$ is the quantum-efficiency and the cross-term $N_{xx}$ is 

\begin{equation}
N_{xx}=\frac{\eta_+ \, \Delta \nu \,QE}{e^{h\,\nu_{CMB}/k_B \,T_{xx}}-1}
\,\;,
\label{equation_3050}
\end{equation}

with $T_{xx}=2 \, T_{e}/\sqrt{\Delta \nu \, t}$.

The second term in Eq. (\ref{equation_3048}), $\delta T_{shot}$, is given by

\begin{equation}
\delta T_{shot}=\frac{2\sqrt{2}\sqrt{t\,N_{th_{up}}+N_{DCR}}\, h\nu_{pump}}{G\,k_B\Delta\nu \,t}
\, \;,
\label{equation_3051}
\end{equation}

where $N_{th_{up}}$ is the number of up-converted thermal photons, $N_{DCR}$ is the dark current's number of counts of the photo-detector,\footnote{As reference, for the PGA series of Princeton Lightwave\textregistered  $\,N_{DCR}=75,000$ counts/s.} $\nu_{pump}$ is the frequency of the laser-pump, $t$ is the integration-time and the gain is $G=\eta_+ \nu_{\text{0}}/\nu_{\text{CMB}}$.

\begin{equation}
N_{th_{up}}=\frac{\eta_+ \, \Delta \nu \,QE}{e^{h\,\nu_{CMB}/k_B \,T_{e}}-1}
\, \;,
\label{equation_3052}
\end{equation}

and finally, the third term ($\delta T_{SP}$) in Eq. (\ref{equation_3048}) is given by

\begin{equation}
\delta T_{SP}=\frac{h \nu_{CMB} } {k_B \, ln \left(\frac{h \nu_{CMB}\, \Delta \nu}{\sigma_{SP}+P_{CMB}}+1\right)}-T_{CMB}
\, \;,
\label{equation_3053}
\end{equation}

where $P_{CMB}$ and $T_{CMB}$ are respectively the power and temperature from the source and the standard deviation of the Super-Poisson distribution ($\sigma_{SP}$) is given by

\begin{equation}
\sigma_{SP}=\frac{\sqrt{(h\nu_{CMB})^2\,t \Delta \nu \, N_{CMB}(1+N_{CMB})}}{t}
\, \;,
\label{equation_3054}
\end{equation}

 $N_{CMB}$ being the number of CMB incoming photons according to the Bose-Einstein distribution.

Thus, the sensitivity of WGM resonator-based radiometers depends on the equivalent temperature ($T_e$), the bandwidth of the device ($\Delta\nu$) and the system efficiency ($\eta_+$) for a given integration time ($\Delta t$) and a given DCR ($n_d$). From Eq. (\ref{equation_3019}) it can be seen that $T_e$ depends on the coupling rates ($\gamma$ and $\gamma'$). Actually, Eq. (\ref{equation_3019}) can be rewritten as 

\begin{equation}
T_{\text{sys}}=\frac{\gamma}{\gamma+\gamma'}T_{\text{A}}+\frac{\gamma'}{\gamma+\gamma'}T_{\text{room}}
\;,
\label{equation_3027}
\end{equation}

where $T_A$ is the antenna temperature. 

Nowadays, the WGM resonator experiments are critically coupled ($\gamma$ = $\gamma'$). The application of Eq. (\ref{equation_3027}) to a standard CMB measurement with $T_A\sim 4$ K and working at room temperature $T_{room}\sim 300$ K yields $T_{sys}\sim 150$ K. Here spurious microwave contributions in real CMB measurement are not considered for simplicity.

On the other hand, in the given references the maximum bandwidth of the WGM resonators is in the order of 1-2 MHz, while the record up-conversion efficiency at 10 GHz is $\eta_{+}\sim 10^{-3}$.

Note that the combination of equations (\ref{equation_3049}) and (\ref{equation_3050}) yield the well known \textit{ideal radiometer equation} applicable to a (classic and) low-frequency regime, which is derived from a quantum context in this case.

Since the bandwidth ($\Delta \nu$) of these WGM resonators is in the MHz range, this experiment cannot be directly transferred to astronomy applications observing continuum, because the ideal radiometer equation establishes that the sensitivity of a radiometer is directly proportional to $\sqrt{\Delta \nu}$, making them non competitive when compared with the HEMT based radiometers, where the bandwidth is in the GHz range \cite{2012SPIE.8452E..33H}. The wide bandwidth gives much higher sensitivities, with one exception which is the case of microwave spectrometers, where the sensitivity is generally given for sub-bands with a few thousands MHz of bandwidth ({\text e.g.}, \cite{1994ApJ...420..457F}, \cite{2007A&A...464..795H}, \cite{2011JCAP...07..025K}, \cite{2011ApJ...730..138S}). Here the possibility of technological transfer could exist for spectral line measurements. The main advantage of a novel spectrometer based on the WGM resonators technology is that while the low noise amplifiers (LNAs) must be cooled down using cryogenics, radiometers based on WGM resonators would be room temperature devices. This is of special interest at frequencies $\nu >> 20$ GHz, where LNAs with thermal noise temperature around 60 K are difficult to achieve.

\begin{figure}[ht!]
		\includegraphics[width=0.9\textwidth]{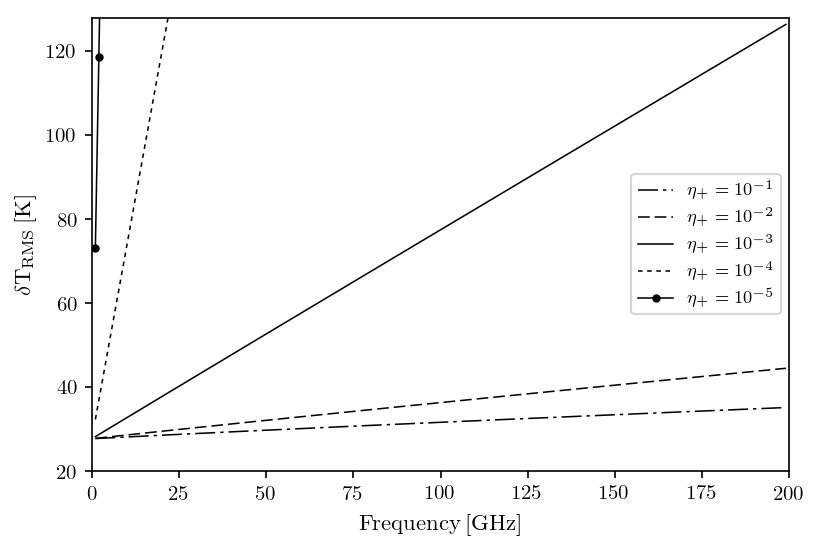}
		\centering 
		\caption{Uncertainty in the estimation of the temperature of the CMB as a function of the conversion efficiency per MHz-bandwidth and ms at room temperature up to 200 GHz.}
		\label{fig3_3}
	\end{figure}

It's interesting now to compare how different values of the up-conversion efficiency ($\eta_+$) affect the uncertainty in the observation with the system proposed in Fig. \ref{fig3_2}. This is shown in Fig. \ref{fig3_3}, where the sensitivity of the system is calculated using (\ref{equation_3048}) and the radiometer is working at room temperature for CMB observations. The  global efficiency of the system has been set at  20\% and the laser-pump is at 1550 nm wavelength. The DCR is 75 kHz. As a reference, the $\delta T_{RMS}$ of the ideal radiometer is around 1 K per MHz-bandwidth and ms, so this novel system will be around two orders of magnitude less sensitive.

A comparative calculation of the theoretical sensitivity of a radiometer based on WGM resonator technology is given in the table \ref{table_3.1}. Here, a regular case of a Dicke's switching \cite{Dicke} cooled HEMT working at 20 GHz and a MKID working at 150 GHz radiometer are compared with six hypothetical cases of radiometers based on the scheme in Fig. \ref{fig3_2}. The sensitivity of the HEMT has been calculated by using the \textit{Dicke's Radiometer Equation} and the sensitivity of the MKID has been extracted from references \cite{Catalano} and \cite{Naruse}, while the sensitivity of the WGM-radiometers has been calculated from Eq. (\ref{equation_3048}).
The WGM1 radiometer works at room temperature under state-of-the-art (and so, realistic) conditions, given by the experiment in \cite{Rueda:16}. The WGM2 works at room temperature with theoretical conversion efficiency $\eta_{+}\sim 10^{-2}$, significantly increases the sensitivity.\footnote{Notice that these are theoretical sensitivities, and in the case of the WGM based radiometers this technology does not exist today.} The WGM3 radiometer is the same WGM2 working at a liquid nitrogen cryogenic temperature around 77 K. Liquid nitrogen cryostats are simpler and easier to maintain and fabricate than helium-cycle cryostats, so this comparison is of interest. The WGM4 is a helium-cycle cryogenic device with $\eta_{+}\sim 10^{-1}$ in order to achieve a high sensitivity, and finally WGM5 and WGM6 are over-coupling setups in order to reduce the thermal noise contribution, which is additionally decreased in WGM6 with cryogenics. The discussion of these results will be included in section \ref{sec:Discussion}.

\begin{table*}
\centering
\begin{tabular}{{cccccccc}}  
\toprule

   & HEMT & WGM1  & WGM2 & WGM3 & WGM4 & WGM5 & WGM6 \\
\midrule

$T_\text{room}$[K]    & - & 300 & 300 & 77 & 14 & 300 & 90\\
$T_\text{sys}$[K]     & 22 & 150 & 150 & 40 & 9 & 60 & 21\\
$\eta_\text{q}$     & - & 0.2  & 0.2 & 0.2 & 0.2 & 0.2 & 0.2 \\
$\eta_{+}$      & - & $10^{-3}$ & $10^{-2}$ & $10^{-2}$ & $10^{-1}$ & $10^{-1}$ & $10^{-1}$  \\
Coupling    & - & critic.  & critic. & critic. & critic. & overc. & overc. \\
Sens[mK$\mathrm{s^{1/2}}$]    & 31 & 1190 & 470 & 195 & 35 & 150 & 65\\
\midrule

   & MKID &   &  &  &  &  &  \\
\midrule

$T_\text{sys}$[K]     & 0.1 &  &  &  &  &  & \\
Sens[mK$\mathrm{s^{1/2}}$]    & 27 & 5850 & 980 & 660 & 79 & 220 & 115\\

\bottomrule

\bottomrule
\end{tabular}
\caption{Comparative of theoretical temperature sensitivities of HEMT and WGM resonator based radiometers at 20 GHz and MKID and WGM resonator based radiometers at 150 GHz per ms and MHz-bandwidth. Duplicated information is omitted in MKID comparison.}
\label{table_3.1}
\end{table*}

\section{A Microwave Spectro-polarimeter Based on WGM Resonators}

\label{sec:Radiometer}
\begin{figure*}[ht!]
\centering
		\includegraphics[width=1\textwidth]{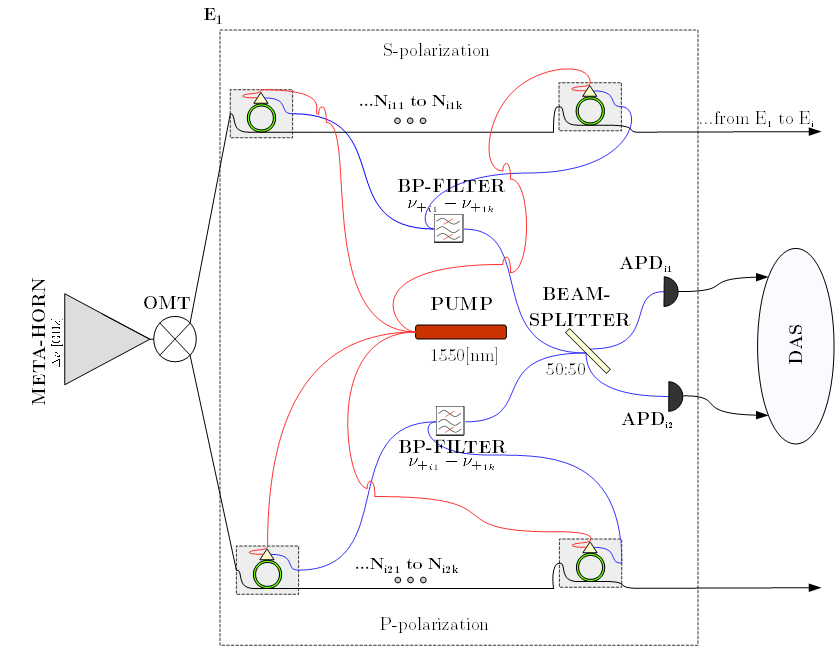}
		\centering 
		\caption{Schematic of a spectro-polarimeter based on WGM resonators technology.}
		\label{fig3_4}
	\end{figure*}

This section presents a preliminary proposal of spectro-polarimeter based on the detector unit presented in Fig. \ref{fig3_2}, using the novel WGM resonator technology. Note that, again, here relative calibration against well-known sources is assumed. In other case, another twin system pointing to a stable load would be necessary. Furthermore, note that our idea is not to present a manufacturable spectro-polarimiter at this point, something that is not feasible with the current technology of WGM resonators, but to imagine how such system would be in a hypothetical future where all the technological limitations of WGM resonators were surpassed.\newline
Fig. \ref{fig3_4} shows the schematic of a spectro-polarimeter for CMB observations thought in order to overcome the actual limitations of the WGM resonators ({\it i.e.}, principally the small bandwidths and low up-conversion efficiency), as discussed in the previous section. Here, a wide band meta-horn (\cite{Miguel_Hern_ndez_2019}, \cite{DEMIGUELHERNANDEZ2019103195}) receives the CMB from the sky and it is separated in to two chains with orthogonal polarization through an orthomode transducer (OMT) before being sent to the $i$-element chain, sharing the same microwave coupling guide, and denoted by $E_i$. The $N_{ijk}$ devices inside an $E_i$ detection element form a filter-bank, because they have been designed to be resonant at different sub-bands. Thus, inside each $E_i$ element the signal covers a sub-band with resonators of bandwidth $\Delta \nu_{r}$. In order to maintain phase coherence of all the $N_{ijk}$ up-converted outputs, the same pump (solid-state laser) is collimated into $N_{ijk}$ mono-mode optic-fibers (OF) or polarization-maintaining optical fiber\footnote{Commonly referred as PANDA.} (PMF or PM fiber) which are coupled into the resonator prisms (color red). The OFs outputs (blue) are merged and filtered in order to select the $\nu_+$ contributions of each resonator ({\it i.e.}, the pass-band is $\nu_{+ij1}-\frac{1}{2}FWHM_{ij1}$ to $\nu_{+ijk}+\frac{1}{2}FWHM_{ijk}$[Hz]) before being forced to pass through a beam-splitter, the outputs f which are sent to twin APDs. Ideally (no systematics, no losses, ...), it is possible to mitigate the thermal-offset and 1/f noise by pseudo-correlation technique. A data acquisition system (DAS)  acquires the signal from all the $E_i$ elements of detector unit. All the $E_i$ elements are similar. All the APD outputs are sent to the same DAS.

In this scheme, each $E_i$ cell contains an array of WGM resonators covering a sub-band given by the desired spectral resolution ($R$), being the numerator of the $N_{ijk}$ resonators and given by

\begin{equation}
i=1, 2, ...,\frac{\Delta \nu}{R}
\;
\label{equation_3045}
\end{equation}

\begin{equation}
k=1,2, ..., \frac{\Delta \nu}{\Delta \nu_{r}}
\;,
\label{equation_3046}
\end{equation}

where $j=1,2$.

\section{Discussion and Future Lines}

\label{sec:Discussion}

In this paper, we have demonstrated the sensitivity of a WGM based radiometer, founding that this technology is not competitive at this point. In this section the discussion of the results and prototypes described in this article is included.\newline
From table \ref{table_3.1}, it is seen that in order to have a WGM radiometer with high sensitivity, it is necessary to reach the value $\eta_{+}\sim 10^{-1}$ and to over-couple or to cool down the system using cryogenics. This up-conversion efficiency is not achievable today, the record efficiency being around $\eta_{+}\sim 10^{-5}$ for the classic double-sideband setup \cite{Sedlmeir:16} and $\eta_{+}\sim 10^{-3}$ for the asymmetric single-sideband experiment in \cite{Rueda:16}. Note that this values of $\eta_{+}$ are reported in experiments, so they include systematics and coupling, uniformity and temperature effects. It is also important to note that the comparison in table \ref{table_3.1} is for 1 MHz bandwidths. The HEMT and MKID radiometers have GHz-bandwidths which leads to much higher sensitivities if a continuum is being measured. 

In order to transfer this technology to radio astronomical applications at room temperature it will be first necessary to increase the bandwidth significantly and also the conversion efficiency of the WGM resonators.

It is important to note that the comparison with HEMP radiometers has been established at microwave frequencies of tens of GHz, where the novel system based on WGM resonators is less competitive. At higher frequencies, where the energy gap between the laser-pumped frequency and the source frequency is lower, the system conversion efficiency is theoretically easier to be maintained in the hundreds of MHz bandwidth, while the efficiency of the LNAs decrease at higher frequencies. However, the LNAs are still more sensitive due to their larger bandwidth. On the other hand, other technologies such as MKID radiometers are more sensitive for higher frequency applications than WGM resonators, something that we pointed in table \ref{table_3.1}. However, MKIDs cannot work under the presence of magnetic fields, so WGMs could be a solution to the question in \cite{madmax} trying to identify a technology that can work in the presence of external magnetic fields over 100 GHz for the detection of axions, although with the current state of the art of the WGM detectors technology the instruments would be too far from the quantum limit.

On the other hand, the technological effort needed to manufacture a competitive prototype based on the diagram in Fig. \ref{fig3_4} is too high at present, since the bandwidth of the WGM resonators is only a few MHz, especially if we take into account that the technology based on HEMTs and MKIDs is very well established and yields more sensitive instruments. 

As final conclusion, WGM technology could be promising for applications in cosmology and radioastronomy if an effort in R\&D is made to achieve a conversion efficiency $\eta_{+}\sim 10^{-1}$, allowing the sensitivities of HEMT and MKID to be achieved, thus being able to benefit from the possibility of working at room temperature, without the need for vacuum and cryogenics.

\section*{Acknowledgements}
The authors would like to thank Luis Enrique Garc\'ia Mu\~noz, Gabriel Santamar\'ia-Botello and Kerlos Atia-Abdalmalak for their input  in the discussion on radiometry and WGM resonators.

\appendix
\label{Appendix}
\section{}
In this section, the fundamentals of nonlinear optical WGM resonators are qualitatively explained. References \cite{mandel_wolf_1995} and \cite{Boyd} will be followed in order to introduce the general theory on nonlinear optics in cavities, and the references \cite{Florian}, \cite{Matsko:07} and \cite{PhysRevA.77.043812} will be followed to adapt the theory to the particular case of a WGM resonator. 

Let's start with the well known definition of density of energy ($w$) in a dielectric 

\begin{equation}
w=\int_{0}^{D} E dD  \;,
\label{equation_3001}
\end{equation}

where $E$ is the electric field and $D$ is the electric displacement. The displacement can be expressed as a function of the polarization $P$ by the relation $D=\varepsilon_0 E+ P$. In the case of a small non-linearity $P$ can be expanded into power series of $E$. This is

\begin{equation}
P_i=\varepsilon_0\left(\sum_{ij}^{} \chi_{ij}^{(1)}E_j+\sum_{ijk}^{} \chi_{ijk}^{(2)}E_jE_k+... \right) \;,
\label{equation_3002}
\end{equation}

where $\chi_{}^{(2)}$ is the symmetry tensor of the material. Since in most cases the polarization exists along or perpendicularly to the optic axis, a unique symmetry tensor is needed. \newline

From (\ref{equation_3001}) and (\ref{equation_3002}) the following expression of the energy in the field can be obtained

\begin{equation}
H=\int w\,dV = \underbrace{\frac{\varepsilon_0}{2} \int E(E+\chi^{(1)}E)\, dV}_{H_0} + \underbrace{\frac{\varepsilon_0}{3} \int \chi^{(2)}E^3\,dV}_{H_{int}}
\;.
\label{equation_3003}
\end{equation}

In (\ref{equation_3003}) $H_0$ represents the undisturbed part of $H$, while $H_{int}$ is the interactive part.\newline

The quantization of the field yields

\begin{equation}
\hat{E}_i=\sqrt{\frac{\hbar \omega_i}{2\varepsilon V_i}}\Psi_i \hat{a}_ie^{-iw_it}+h.c.
\;,
\label{equation_3004}
\end{equation}

where the terms $\Psi_i$ is the spatial modal distribution which contains the information of the phase, the normalization volume is $V_i=\int_{V}\Psi_i \Psi_i^{*}$ and $\omega_i$ contains the frequency information.The permittivity is related with the refractive index $n_i$ by $\varepsilon_i=\varepsilon_0n_i^2$ and $\hat{a}_i$ is the annihilation operator.\newline

Inserting (\ref{equation_3004}) into the $H_{int}$ part of (\ref{equation_3003}) results in the expression of the interaction Hamiltonian

\begin{equation}
\hat{H}_{\text{int}}=\frac{\varepsilon_0}{3}\chi^{(2)}\left(\frac{\hbar}{2}\right)^{\frac{3}{2}}\int \left(\sum_{i}\Psi_i \sqrt{\frac{ \omega_i}{\varepsilon V_i}} \hat{a}_ie^{-iw_nt}+h.c.\right)^3dV
\;.
\label{equation_3005}
\end{equation}  \newline

The interaction Hamiltonian in (\ref{equation_3005}) is in general a sum of $4i^3$ terms and their hermitian conjugates ($h.c.$), and represents a nonlinear process. The sum-frequency generation (SFG), where two photons of arbitrary modes 1 and 2 are annihilated creating a photon in a mode 3 (triplet $\hat{a_1}\hat{a_2}\hat{a_3}^{\dagger}$) and its inverse, the difference-frequency generation (DFG) ($\hat{a_1}^{\dagger}\hat{a_2}^{\dagger}\hat{a_3}$) or parametric down conversion (PDC) can be examined for the case of 3 different frequencies ($\omega_1,\omega_2,\omega_3$) and their corresponding operators ($\hat{a_1},\hat{a_2},\hat{a_3}$). Thus, expanding (\ref{equation_3005}) and limiting the expansion to the six relevant terms yields

\begin{equation}
\begin{split}
\hat{H}_{\text{int}}=2\varepsilon_0\chi^{(2)}\left(\frac{\hbar}{2}\right)^{\frac{3}{2}}\sqrt{\frac{\omega_1\omega_2\omega_3}{\varepsilon_1\varepsilon_2\varepsilon_3V_1V_2V_3}}   \\ \cdot \int  \Psi_1\Psi_2\Psi_3^{*}\hat{a_1}\hat{a_2}\hat{a_3}^{\dagger}e^{-i(\omega_1+\omega_2-\omega_3)t}dV+h.c.
\;.
\label{equation_3006}
\end{split}
\end{equation}  \newline

In the case of net interaction, the time average of (\ref{equation_3006}) tends to nonzero. The time dependent part of (\ref{equation_3006}) tends to zero when

\begin{equation}
\omega_3= \omega_1+\omega_2
\;.
\label{equation_3007}
\end{equation}  

Reference \cite{Manley} shows that the modes can only be exchanged in multiples of $\hbar \omega$, so (\ref{equation_3007}) can be interpreted as a conservation of the energy through the interaction. \newline
In the case of WGM resonators, the spatial azimuthal part can be separated from the radial and polar parts

\begin{equation}
\Psi_{mpq}(r)=\psi_{mpq}(r,\theta)e^{-im\phi}
\;,
\label{equation_3008}
\end{equation}  

where $m$ is the azimuthal mode number describing the field equatorial plane oscillations and the associated angular momentum of the mode, $p$ is the polar mode number describing the field minima in the polar direction and $q$ is the radial mode number which describes the number of maxima in this direction. Modes with $q = 1$ and $p = 0$ are called \textit{fundamentals}. \newline

Inserting (\ref{equation_3008}) into (\ref{equation_3006}) reveals that only when the following constraint is satisfied the volumetric integral is non-zero 

\begin{equation}
m_3=m_1+m_2
\;.
\label{equation_3009}
\end{equation}  \newline

Equation (\ref{equation_3009}) expresses the conservation of angular momentum in a WGM resonator. Since the propagation constant in a WGM resonator is $\beta=m/R=k n_{e}$, where $R$ is the the resonator radius and $n_{e}$ is the effective refractive index. This equation can be related with a phase matching condition for plane waves with conservation of the wavenumber, {\it i.e.}, $n_1k_1+n_2k_2=n_3k_3$, the difference being that $n_{e}$ depends on the geometry of the resonator. From (\ref{equation_3006}) and following \cite{PhysRevA.77.043817}, two more mode selection rules can be derived

\begin{equation}
|m_1-m_2+p_2-p_1| \le m_3-p_3 \le m_1+m_2-p_1-p_2
\;,
\label{equation_3010}
\end{equation}  

and

\begin{equation}
m_1+m_2+m_3-p_1-p_2-p_3 \in 2\mathds{Z}
\;.
\label{equation_3011}
\end{equation} 

The Eq. (\ref{equation_3010}) and (\ref{equation_3011}) were originally derived for spherical resonators, but because of the polar symmetry they can be applied to the case of WGM resonators too. For the particular case of $p=0$, both equations are satisfied if (\ref{equation_3009}) is also kept. \newline

It can be shown that when phase-matching and energy conservation exists, (\ref{equation_3006}) simplifies and the nonlinear coupling factor can be derived

\begin{equation}
g=2\varepsilon_0\chi^{(2)}\sqrt{\frac{\hbar}{2}}\sqrt{\frac{\omega_1\omega_2\omega_2}{\varepsilon_1\varepsilon_2\varepsilon_3V_1V_2V_3}}\pi R\int_{S}\psi_1\psi_2\psi_3\,dA
\;,
\label{equation_3012}
\end{equation} 

where $R$ is the radius of the resonator and the volumetric integral becomes a surface integral due to the fact that the azimuthal dependence vanishes because of the phase-matching. On the other hand, the normalization volumes also become surfaces

\begin{equation}
V_i=\int\Psi_i\Psi_i^*\,dV=2\pi R\int\psi_i^2\,dA
\;.
\label{equation_3013}
\end{equation}

From (\ref{equation_3012}) and (\ref{equation_3013}) is possible to obtain

\begin{equation}
g= \frac{1}{2} \sqrt{\frac{\hbar}{\pi R}}\epsilon_0 \chi^{(2)} \sqrt{\frac{\omega_1 \omega_2 \omega_3}{\epsilon_1 \epsilon_2 \epsilon_3}}
\frac{\int_{S}\psi_1\psi_{2}\psi_3}{\sqrt{\int_S\psi_1^2\times\int_S\psi_2^2\times\int_S\psi_3^2}}
\;.
\label{equation_3014}
\end{equation} \newline

The analysis of (\ref{equation_3014}) yields several conclusions. In the first place, the nonlinear coupling factor $g$ scales with the nonlinear polarity $\chi^{(2)}$. Secondly, $g$ scales inversely with $\sqrt{R}$ and $\Psi \propto \sqrt{\mathcal{F}} \propto 1/\sqrt{R}$, where the finnese is given by

\begin{equation}
\mathcal{F}=\frac{2\gamma}{(\gamma ' +\gamma)^2} \frac {1}{\tau}
\;,
\label{equation_3015}
\end{equation}

where the free spectral range (FSR) is $1/\tau$, where $\tau \simeq 2 \pi R n/c$, being $R$ the resonator radius, $n$ the refraction index and $c$ the speed of light, the {\it round trip time}. Thirdly the nonlinear conversion process is better for small cross-sections of field distribution. The coupling rate is given by

\begin{equation} 
\gamma=\frac{| \mathfrak{t} |^2}{2\tau}
\;,
\label{equation_A3016}
\end{equation}

with $\mathfrak{t}\in \mathds{C}$ being the transmission coefficient, which can be derived from the Fresnel's equations, and the field loss rate is

\begin{equation}
\gamma'=\frac{1-a}{\tau}
\;,
\label{equation_A3017}
\end{equation} \newline

where a is the attenuation factor. The situation $\gamma<\gamma'$ is called the \textit{undercoupling} case, $\gamma>\gamma'$ \textit{overcoupling} and $\gamma = \gamma '$ is the \textit{critical coupling} case. Note that Eqs. (\ref{equation_A3016}) and (\ref{equation_A3017}) are the same Eqs. (\ref{equation_3016}) and (\ref{equation_3017}).  \newline

The solution for the rate equations given in appendix C \cite{Florian}  for a WGM resonator makes it possible to obtain the photon occupation number

\begin{equation}
\langle n_{{th}}\rangle=\frac{\gamma \, n_{th_{\text{s}}}}{\gamma ' +\gamma}+\frac{\gamma' n_{th_{\text{i}}}}{\gamma ' +\gamma}
\;,
\label{equation_3018}
\end{equation} 
where addends for both the surrounding (s) medium and the resonator material (i) are included.

When the environment and the resonator have different temperatures, an effective temperature ($T_{\text{e}}$) is achieved in the thermal occupation of the resonator modes

\begin{equation}
T_{\text{e}}=\frac{\gamma}{\gamma+\gamma'}T_{\text{s}}+\frac{\gamma'}{\gamma+\gamma'}T_{\text{i}}
\;.
\label{equation_A3019}
\end{equation}

From (\ref{equation_A3019}) it can derived that if the environment is {\it cold} and and the resonator is over-coupled, $T_{\text{e}}$ can be decreased. This property is relevant for astronomical purposes, because it implies that the thermal noise can be naturally decreased. Note that Eq. (\ref{equation_A3019}) is the same Eq. (\ref{equation_3019}). \newline

In a WGM resonator, the power of the up-converted signal scales linearly with the power of the microwave input power. Thus, it is possible to indirectly measure a microwave incoming signal from a source once converted to the the optical range. Furthermore, this conversion process is coherent so it conserves the phase information. Fig.\ref{fig3_1} represents the principle of the up-conversion, where $\Omega=\omega-\omega_-$ is the microwave pulse, $\omega_0$ is the optical pump pulse and the converted frequency is given by $\omega_+=\omega+\Omega$. The factor $m$ is an integer.

The Hamiltonian which describes the process represented in Fig.\ref{fig3_1} is

\begin{equation}
\begin{split}
\hat{H}=\underbrace{\hbar \omega_0 \hat{a}^{\dagger}\hat{a}+\hbar \omega_- \hat{b}^{\dagger}_- \hat{b}_-+\hbar \omega_+ \hat{b}^{\dagger}_+ \hat{b}_++\hbar \Omega \hat{c}^{\dagger}\hat{c}}_{H_0}+ \\ \hbar g (\underbrace{ \hat{a} \hat{b}^{\dagger}_- \hat{c}^{\dagger}+\hat{a}^{\dagger}\hat{b}_- \hat{c}}_{DFG}+\underbrace{ \hat{a}\hat{b}^{\dagger}_+\hat{c}+\hat{a}^{\dagger}\hat{b}_+ \hat{c}^{\dagger}}_{SFG})
\;,
\label{equation_3020}
\end{split}
\end{equation}

where $\hat{a}$ is associated with the mode of the laser-pump, $\hat{b}_-$ with the lower frequency detuned sideband, $\hat{b}_+$ with the higher frequency detuned sideband and $\hat{c}$ to the microwave incoming signal. Notice that the $g$ coupling factor term could be different for DFG and SFG. The power of the sidebands is given by

\begin{equation}
P_{\pm}=8\frac{g^2}{\hbar}\frac{\gamma^2\gamma_{\Omega}}{|\Gamma_A|^2 |\Gamma_B|^2 |\Gamma_C|^2} \frac{\omega_B}{\omega_A \Omega}P_AP_{\Omega}
\;,
\label{equation_3021}
\end{equation}

where $\Omega$ is the microwave frequency, and

\begin{equation}
\Gamma_A=\gamma+\gamma'+i(\omega_0-\omega)
\;
\label{equation_3022}
\end{equation}

\begin{equation}
\Gamma_{B_-}=\gamma+\gamma'+i[\omega_--(\omega-\Omega)]
\;
\label{equation_3023}
\end{equation}

\begin{equation}
\Gamma_{B_+}=\gamma+\gamma'+i[\omega_+-(\omega+\Omega)]
\;
\label{equation_3024}
\end{equation}

\begin{equation}
\Gamma_{C}=\gamma_{\Omega}+\gamma_{\Omega}'+i(\Omega_0-\Omega)
\;,
\label{equation_3025}
\end{equation}

where $A$, $B_{\pm}$, and $C$ are the slowly-varying amplitudes of the operators $\hat{a}$, $\hat{b}$, and $\hat{c}$ respectively (e.g., see \cite{Matsko:07} Eq. 41 in section 3 to know more). From (\ref{equation_3021}) it can be deduced that the power of the sidebands depends linearly on the power of the optical pump and on the power of the microwave incoming signal. The conversion presents quadratic dependence on the coupling factor ($g$) and the optical quality factor ($Q$). In order to obtain an efficient conversion, these parameters should be optimized.

In the system in Fig.\ref{fig3_1}, assuming that the microwave modes occupy larger volumes than the optical modes, the nonlinear coupling factor becomes

\begin{equation}
g=2\chi^{(2)}\frac{\omega}{n^2}\sqrt{\frac{\hbar \Omega}{2\varepsilon_{\Omega}V_{\Omega}}}\Psi_{\Omega} (r_0)
\;,
\label{equation_3026}
\end{equation}

where $\Psi_{\Omega} (r_0)$ is the microwave field location dependent amplitude function and $n$ is the modal photon number.


\bibliography{mybibfile}

\end{document}